\documentclass{article}

\usepackage{arxiv}

\usepackage[utf8]{inputenc} % allow utf-8 input
\usepackage[T1]{fontenc}    % use 8-bit T1 fonts
\usepackage{hyperref}       % hyperlinks
\usepackage{url}            % simple URL typesetting
\usepackage{booktabs}       % professional-quality tables
\usepackage{amsfonts}       % blackboard math symbols
\usepackage{nicefrac}       % compact symbols for 1/2, etc.
\usepackage{microtype}      % microtypography
\usepackage{lipsum}		% Can be removed after putting your text content
\usepackage{graphicx}
\usepackage{tikz}
\usepackage{subfig}
\usetikzlibrary{positioning, calc}
\usetikzlibrary {arrows.meta, decorations.pathmorphing}

\title{Coupling a Recurrent Neural Network to SPAD TCSPC Systems for Real-time Fluorescence Lifetime Imaging}

\author{ Yang Lin \quad Paul Mos \quad Andrei Ardelean \quad Claudio Bruschini \quad Edoardo Charbon \\
	Advanced Quantum Architecture Laboratory\\
	École polytechnique fédérale de Lausanne\\
	Neuchâtel, 2002, Switzerland \\
	\texttt{\{yang.lin, paul.mos, andrei.ardelean, claudio.bruschini, edoardo.charbon\}@epfl.ch} \\
}

\renewcommand{\shorttitle}{RNN-coupled SPAD TCSPC System for Real-time FLI}

\begin{document}
\maketitle

\begin{abstract}
Fluorescence lifetime imaging (FLI) has been receiving increased attention in recent years as a powerful diagnostic technique in biological and medical research. However, existing FLI systems often suffer from a tradeoff between processing speed, accuracy, and robustness. In this paper, we propose a robust approach that enables fast FLI with no degradation of accuracy. The approach is based on a SPAD TCSPC system coupled to a recurrent neural network (RNN) that accurately estimates the fluorescence lifetime directly from raw timestamps without building histograms, thereby drastically reducing transfer data volumes and hardware resource utilization, thus enabling FLI acquisition at video rate. We train two variants of the RNN on a synthetic dataset and compare the results to those obtained using center-of-mass method (CMM) and least squares fitting (LS fitting). Results demonstrate that two RNN variants, gated recurrent unit (GRU) and long short-term memory (LSTM), are comparable to CMM and LS fitting in terms of accuracy, while outperforming them in background noise by a large margin. 
To explore the ultimate limits of the approach, we derived the Cramer-Rao lower bound of the measurement, showing that RNN yields lifetime estimations with near-optimal precision. Moreover, our FLI model, which is purely trained on synthetic datasets, works well with never-seen-before, real-world data. To demonstrate real-time operation, we have built a FLI microscope based on Piccolo, a 32x32 SPAD sensor developed in our lab. Four quantized GRU cores, capable of processing up to 4 million photons per second, are deployed on a Xilinx Kintex-7 FPGA. Powered by the GRU, the FLI setup can retrieve real-time fluorescence lifetime images at up to 10 frames per second. The proposed FLI system is promising and ideally suited for biomedical applications, including biological imaging, biomedical diagnostics, and fluorescence-assisted surgery, etc.
\end{abstract}

% keywords can be removed
\keywords{FLIM \and SPAD \and Neural network}

\section*{Introduction}

% introduction to FLI
% application
% challenges posed by new applications

Fluorescence lifetime imaging (FLI) is an imaging technique for the characterization of molecules based on time they decay from an excited state to the ground state \cite{Datta.2020}. Compared with fluorescence intensity imaging, FLI is insensitive to excitation intensity fluctuations, variable probe concentration, and limited photobleaching. Besides, through the appropriate use of targeted fluorophores, FLI is able to quantitatively measure the parameters of the microenvironment around fluorescent molecules, such as pH, viscosity, and ion concentrations\cite{vanMunster., Suhling.2015}. With these advantages, FLI has wide applications in the biological sciences, for example to monitor protein-protein interactions\cite{Wallrabe.2005}, and plays an increasing role in medical and clinical settings such as visualization of tumor margins\cite{Unger.2020}, cancerous tissue detection\cite{Datta.2020,Erkkila.2020}, and computer-assisted robotic surgery\cite{Weyers.2019,Phipps.}.

% Acquiring an accurate lifetime image in a short time, however, still remains a challenge. Fast FLI systems are required in biomedical and clinical scenarios such as the study of fast-moving cells and cancerous tissue detection\cite{Bruza.2021, Sun.2013, AlfonsoGarcia.2021}. Most high-accuracy FLI systems detect photons with single-photon detectors, time-tag incoming photons, and send the timestamps to the PC for processing in a technique known as time-correlated single-photon counting (TCSPC). This technique lays a great burden on data transfer and data storage. Algorithms to determine lifetime from timestamps often suffer from low photon count, background noise, dark count, saturation, and short acquisition time.

\begin{figure}
    \centering
    \includegraphics[width=\textwidth]{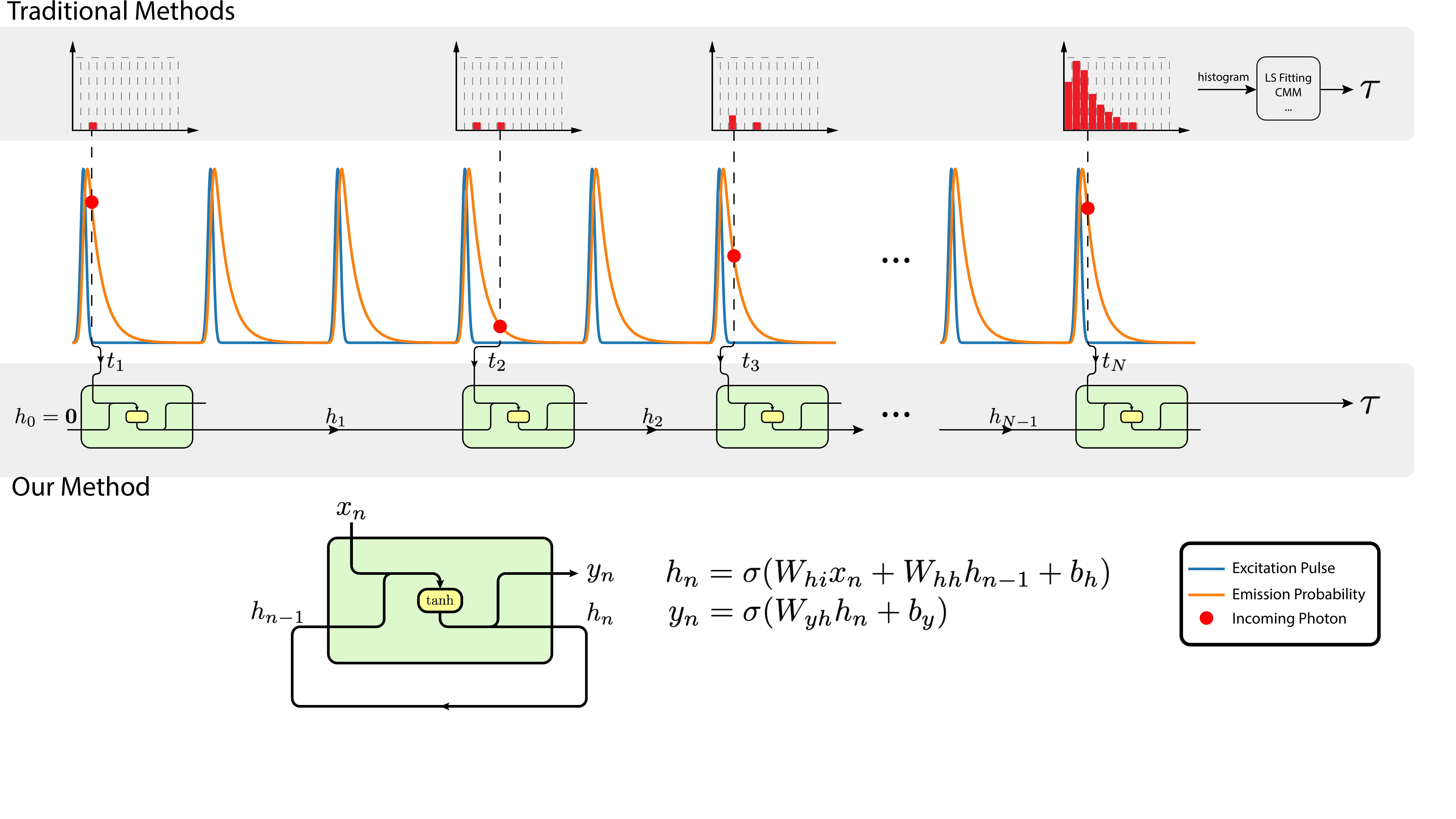}
    \caption{In a traditional TCSPC FLI system, the sample is excited by a laser repeatedly, and the emission photons are detected and time-tagged. A histogram is gradually built on these timestamps, from which the lifetime can be extracted after the acquisition is completed. In our proposed system, upon the receiving of a photon, the timestamp is fed into the RNN immediately. The RNN updates the hidden state accordingly and idles for the next photon. The schematic and formula of simple RNN are shown here. At timestep $n$, the RNN takes the current information $x_n$ and the past information $h_{n-1}$ as input, then updates the memory to the current information $h_n$ and gives out a prediction $y_n$.}
    \label{fig:tcspc-rnn}
\end{figure}

% instrumentation of FLI
% traditional bulky FLI
% advantage of SPAD 
Time-correlated single-photon counting (TCSPC) is popular among FLI systems due to its superiority over other techniques in terms of time resolution, dynamic range, and robustness. In TCSPC, one records the arrival time of individual photons when emitted by molecules upon photoexcitation \cite{Becker., Becker.2012c, Hirvonen.2017}. After repeated measurements, one can construct a histogram of photon arrivals, which closely matches the true response of molecules, thus enabling the extraction of FLI, as shown Figure~\ref{fig:tcspc-rnn}. The instrumentation of a typical TCPSC FLI system features a confocal setup, including a single-photon detector, a dedicated TCSPC module for time tagging, and a PC for lifetime estimation\cite{Kapusta.2015, Becker.}. Such systems are mostly unsuitable for rising clinical applications such as non-invasive monitoring, where a miniaturized and fast TCSPC system is desired \cite{AlfonsoGarcia.2021}. Besides, the large amount of data generated by TCSPC lays a great burden on data transfer, data storage, and data processing. A powerful PC, sometimes equipped with dedicated GPUs, is required to acquire and process TCSPC data. 
TCSPC requires photodetectors with picosecond time resolution and single-photon detection capability. In the last decade, single-photon avalanche diodes (SPADs) have been used successfully in TCSPC systems and, with the advent of CMOS SPADs, the expansion of these detectors into high-resolution image sensors for widefield imaging has 
been accomplished successfully \cite{Morimoto.2021}. 
Several reviews of the use of SPADs in biophotonics have recently appeared \cite{Caccia.2019, Bruschini.2019, Cusini.2022}.

% classic lifetime determination method
Least-square (LS) fitting and maximum likelihood estimation (MLE) are widely used for fluorescence lifetime estimation\cite{Grinvald.1974, Bajzer.1991, A.Chessel.2013}. These two methods rely on iterations to achieve high accuracy, but they are time-consuming since computationally expensive. Various non-fitting methods have been proposed to tackle these problems but often compromise on other specifications, among which the Center-of-Mass method (CMM) is a typical one. CMM is a simple, fast, and photon-efficient alternative, which has been already applied in some real-time FLI systems\cite{Isenberg.1969, Li.2010b, Li.2011}. However, it is very sensitive to background noise, and the estimation is biased due to the use of truncated histograms\cite{Liu.2019}. 

% Rapid lifetime determination (RLD) also offers simple and fast calculation and shows potential for real-time FLI, but it proves to be less accurate than the LS fitting\cite{Ballew.1989,AhmetT.Erdogan.2019}. Phasor analysis needs more complex computation than CMM and RLD, but is still much faster than fitting methods \cite{Digman.2008}. It requires prior knowledge of the instrument response function and thus careful calibration.

% photobleaching

% existing NN-based FLI
Neural networks provide a new path to fast fluorescence lifetime extraction\cite{Mannam.2020}. The first neural network-based model for fluorescence lifetime estimation was proposed in 2016, where higher accuracy and faster processing than LS fitting were reported\cite{Wu.2016}. Since then, several neural network architectures, including fully connected neural network (FCNN), convolutional neural network (CNN), and generative adversarial network (GAN) solutions have been explored for this end\cite{Smith.2019b, Zickus.2020, Xiao.2021, Zang.3252022, Chen.2022}. These techniques showed the ability to resolve multi-exponential decays and achieve accurate and fast estimation even in low photon-count scenarios. Apart from fluorescence lifetime determination, these neural networks can extract high-level features and can be integrated into a large-scale neural network for end-to-end lifetime image analysis such as cancerous tissue margin detection\cite{Marsden.2021} and microglia detection\cite{Sagar.2020}.
% pitfall of neural network-based FLIM

% our work
In this work, we propose to adopt the paradigm of edge artificial intelligence (Edge AI), constructing a recurrent neural network (RNN) -coupled SPAD TCSPC system for real-time FLI. We train and test variants of RNNs for lifetime estimation and deploy them on FPGA to realize event-driven and near-sensor processing. The working principle is illustrated in Figure~\ref{fig:tcspc-rnn}. Upon the arrival of  photons, the timestamp is processed by the RNN directly without histogramming. From photon detection to lifetime estimation, the whole system is integrated into a miniaturized device, which achieves reduced data transfer rates. With the flexibility to retrain neural networks, the same system can be easily reused for other very different applications, such as classification.

\section*{Results}

The proposed system comprises a SPAD image sensor with timestamping capability coupled to an FPGA for implementation of neural networks {\em in situ}. In this section we describe the utilized RNN, its training, and the achieved results. We also describe the upperbounds on accuracy that were derived to contextualize the results obtained with the RNNs.

\subsection*{RNNs trained on Synthetic Datasets and Performance}

We train and evaluate RNNs on synthetic datasets. Three RNN variants, namely simple RNN, gated recurrent unit (GRU)\cite{cho2014learning}, and long short-term memory (LSTM)\cite{hochreiter1997long}, are adopted. These RNNs are constructed with 8, 16, and 32 hidden units, respectively. LS fitting and CMM are also benchmarked. The metrics used for evaluation are the root mean squared error (RMSE), mean absolute error (MAE), and mean absolute percentage error (MAPE). A common range of lifetime is covered here, which is from 0.2 to 5 ns, and we assume a laser repetition frequency of 20 MHz. %%% ADD REFERENCE to METHODS section.

We start with a simple situation in which background noise is absent. All the tests are run on a PC with 32-bit floating point (32FP) precision. The results are presented in Table~\ref{tab:rnnresults}. One can observe that CMM achieves the lowest error in MAE and MAPE, GRU-32 achieves the lowest RMSE error, and GRU-32 and LSTM-32 have very similar performance to CMM. The performance of CMM itself is understandable. In this case, background noise is not considered, and the repetition period is 10 times the longest lifetime. Under these conditions, CMM is very close to the maximum likelihood estimator of the lifetime. When comparing Simple RNN, GRU, and LSTM, one can observe that GRU outperforms LSTM by a small margin, and both of them perform much better than Simple RNN. As we can see with the decrease in model size, errors increase accordingly.
%%% The later studies show that the difference between the prediction of 32FP neural networks and quantized neural networks is below 0.01\%. [Move to FPGA section.]

\begin{table}[!tbp]
	\centering
	\begin{tabular}{cccc}
		\toprule
		Model	& RMSE	& MAE	& MAPE \\
		\midrule
		LS Fitting & 0.1642 & 0.1201 & 0.0553 \\
		CMM & \textbf{0.0915} & \textbf{0.0642} & \textbf{0.0250} \\
		& & & \\
		Simple RNN-8 & 0.2516 & 0.1979 & 0.0969 \\
		Simple RNN-16 &  0.2396 & 0.1798 & 0.0771 \\
		Simple RNN-32 & 0.1877 & 0.1415 & 0.0659 \\
		GRU-8 & 0.0957 & 0.0695 & 0.0297 \\
		GRU-16 & 0.0928 & 0.0666 & 0.0274 \\
		GRU-32 & \textbf{0.0908} & \textbf{0.0647} & \textbf{0.0261} \\
		LSTM-8 & 0.0981 & 0.0720 & 0.0423 \\
		LSTM-16 & 0.0928 & 0.0669 & 0.0277 \\
		LSTM-32 & 0.0916 & 0.0656 & 0.0267 \\
		\bottomrule
	\end{tabular}
	\caption{RNN models are trained and tested on a synthetic dataset, where the fluorescence decay model is mono-exponential, lifetime ranges from 0.2 and 5 ns, laser repetition frequency is 20 MHz, and background noise is not considered. Their performance is benchmarked against Least-square (LS) fitting and Center-of-Mass method (CMM). RMSE: root mean squared error, MAE: mean absolute error, MAPE: mean absolute percentage error.}
	\label{tab:rnnresults}
\end{table}

Background noise is often inevitable during fluorescence lifetime imaging, especially in diagnostic and clinical setups where the interruption to existing workflows is supposed to be minimized\cite{AlfonsoGarcia.2021}. In our FLI system, it is estimated that at least 1\% of the collected timestamps are from background noise. Therefore, we study the performance of each method under varying background noise levels. For simplicity, only LSTM-32 is used to compare with benchmarks. LSTM-32 is trained on a synthetic dataset, where 0 to 10\% uniform background noise is added to the samples randomly. Here we also illustrate the result of CMM with background noise subtraction, assuming that the number of photons from background noise is known, though it is often not the case in real-time FLI systems. Two synthetic datasets are built for evaluation, where the background noise ratios are 1\% (SNR=20dB) and 5\% (SNR=12.8dB), respectively. The results are presented in Table~\ref{tab:bgnoise}. We can see that LSTM-32 outperforms other methods in all metrics and scenarios. Combined with Table~\ref{tab:rnnresults}, one can observe that errors increase when the background noise increases for all the methods. However, LSTM and LS fitting are more robust to background noise, while CMM is extremely sensitive to it. This finding is in agreement with previous studies\cite{Bouchet.2019,Datta.2020}.

\begin{table}[!tbp]
    \centering
\begin{tabular}{ccccccc}
    \toprule
     &  \multicolumn{3}{c}{1\% Background Noise} & \multicolumn{3}{c}{5\% Background Noise}\\
     & RMSE & MAE & MAPE & RMSE & MAE & MAPE \\
     \midrule
    LS fitting & 0.1678 & 0.1226 & 0.0562 & 0.1883 & 0.1368 & 0.0609 \\
    CMM & 0.2367 & 0.2168 & 0.1577 & 1.0742 & 1.0635 & 0.7799 \\
    CMM* & 0.1099 & 0.0839 & 0.0456 & 0.2476 & 0.2128 & 0.1444 \\
    LSTM-32 & \textbf{0.1019} & \textbf{0.0733} & \textbf{0.0304} & \textbf{0.1097} & \textbf{0.0784} & \textbf{0.0323} \\
%%%    GRU-32 & \textbf{000} & \textbf{0.0733} & \textbf{0.0304} & \textbf{0.1097} & \textbf{0.0784} & \textbf{0.0323} \\
    \bottomrule
\end{tabular}
    \caption{Performance of LS fitting, CMM, CMM with background subtraction, and LSTM-32 in the presence of 1\% and 5\% background noise. *CMM with background noise subtraction. LSTM-32 is trained on a dataset including 0\% to 10\% random background noise for generalization in different scenarios.}%%% [Specify training conditions.]
    \label{tab:bgnoise}
\end{table}

\subsection*{Cramer-Rao Lower Bound Analysis}

\begin{figure}

\begin{minipage}{\textwidth}
    \centering
    \subfloat[Relative standard deviation of different estimation methods and CRLB as a function of lifetime for a fixed number of detected photons (1024). No background noise.\label{fig:crlb-lifetime}]{
        \includegraphics[width=0.45\textwidth]{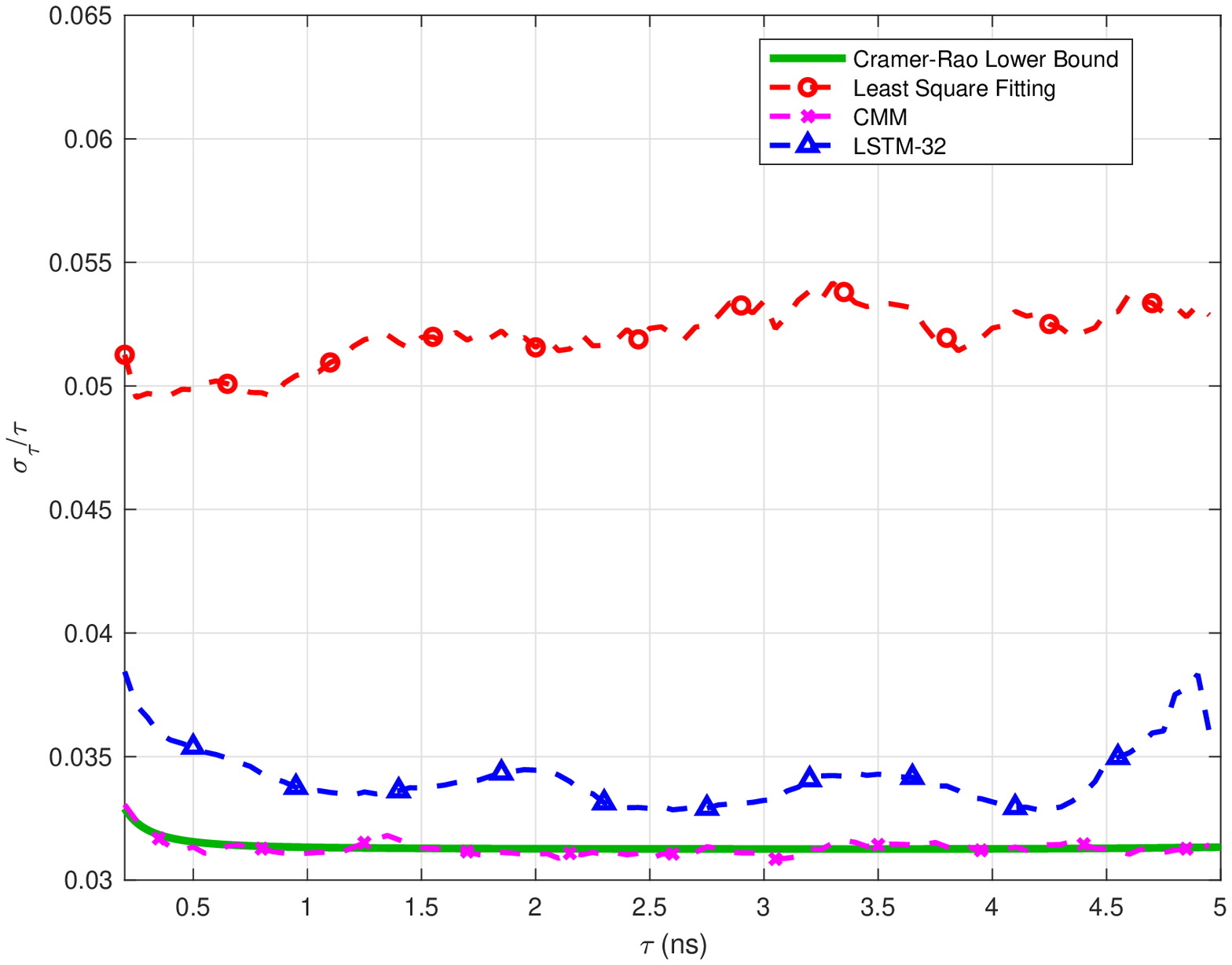}
    }
    \subfloat[Relative standard deviation of different methods and CRLB as a function of the number of detected photons for a fixed lifetime of 2.5 ns. No background noise.\label{fig:crlb-photon}]{
        \includegraphics[width=0.45\textwidth]{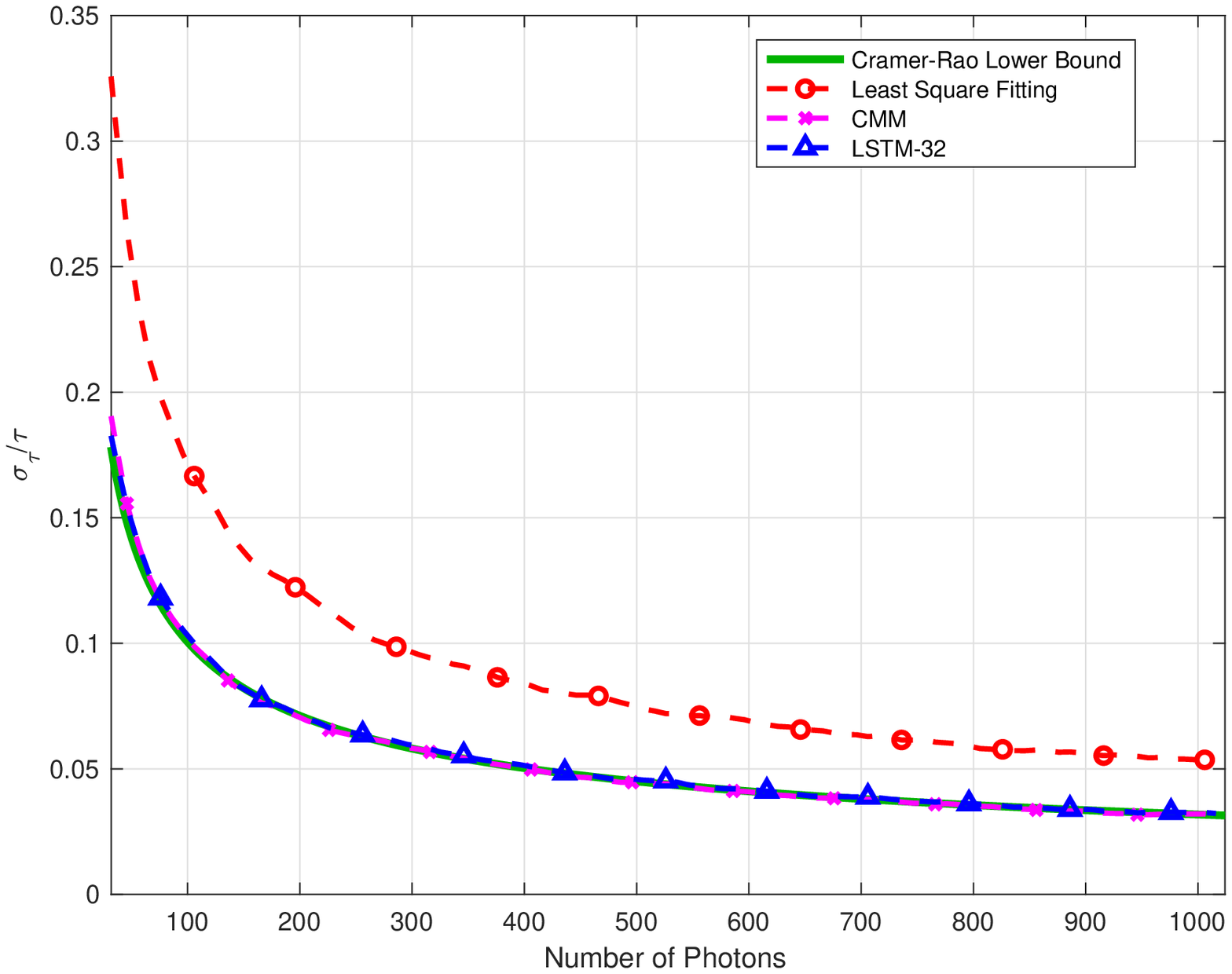}
    }
    % \caption{Cramer-Rao lower bound analysis.}
    % \label{fig:crlb}
\end{minipage}

\begin{minipage}{\textwidth}
    \centering
    \subfloat[Relative standard deviation of different estimation methods and CRLB as a function of lifetime for a fixed number of detected photons (1024). 1\% background noise.\label{fig:crlb-lifetime-1per}]{
        \includegraphics[width=0.45\textwidth]{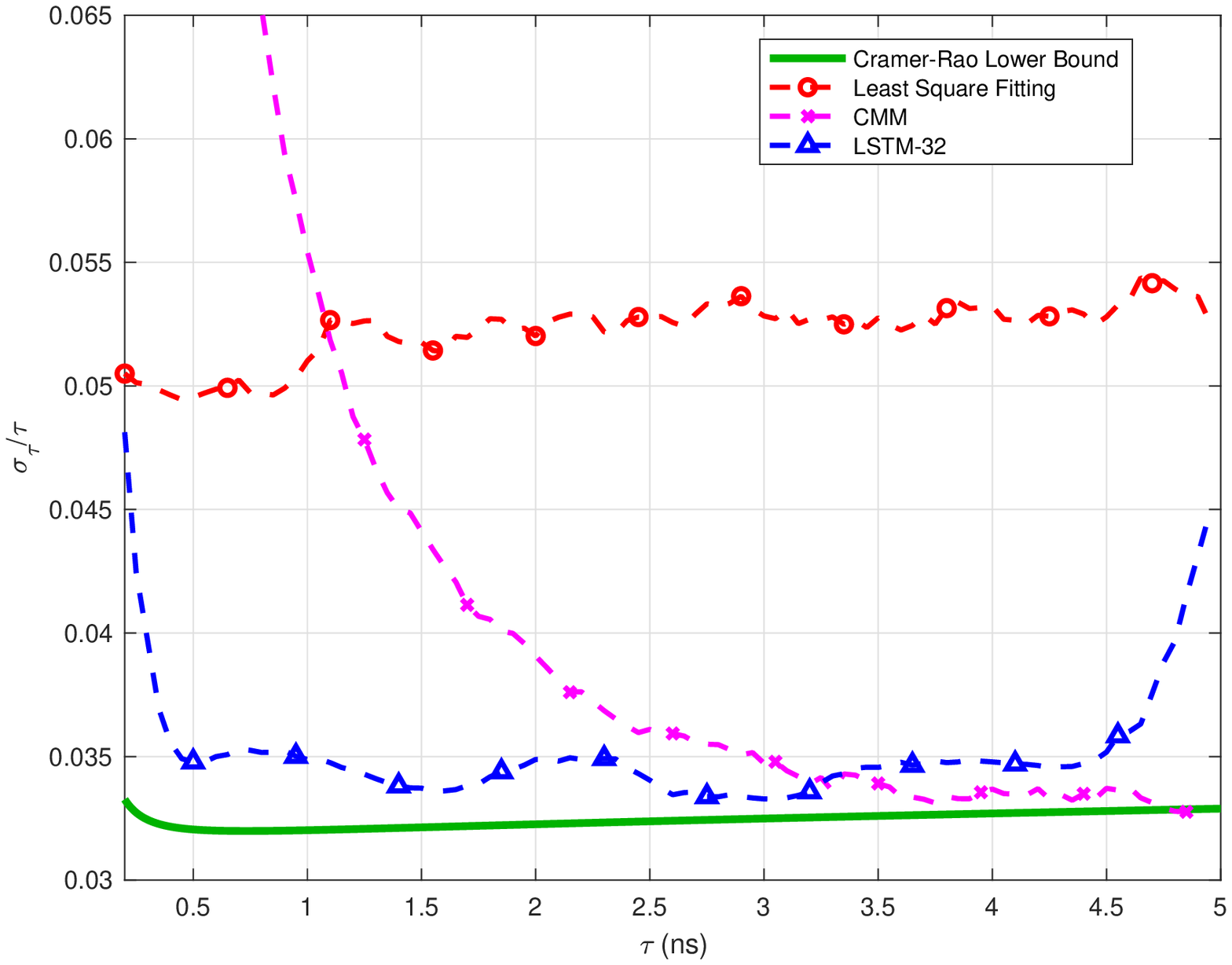}
    }
    \subfloat[Relative standard deviation of different methods and CRLB as a function of the number of detected photons for a fixed lifetime of 2.5 ns. 1\% background noise.\label{fig:crlb-photon-1per}]{
        \includegraphics[width=0.45\textwidth]{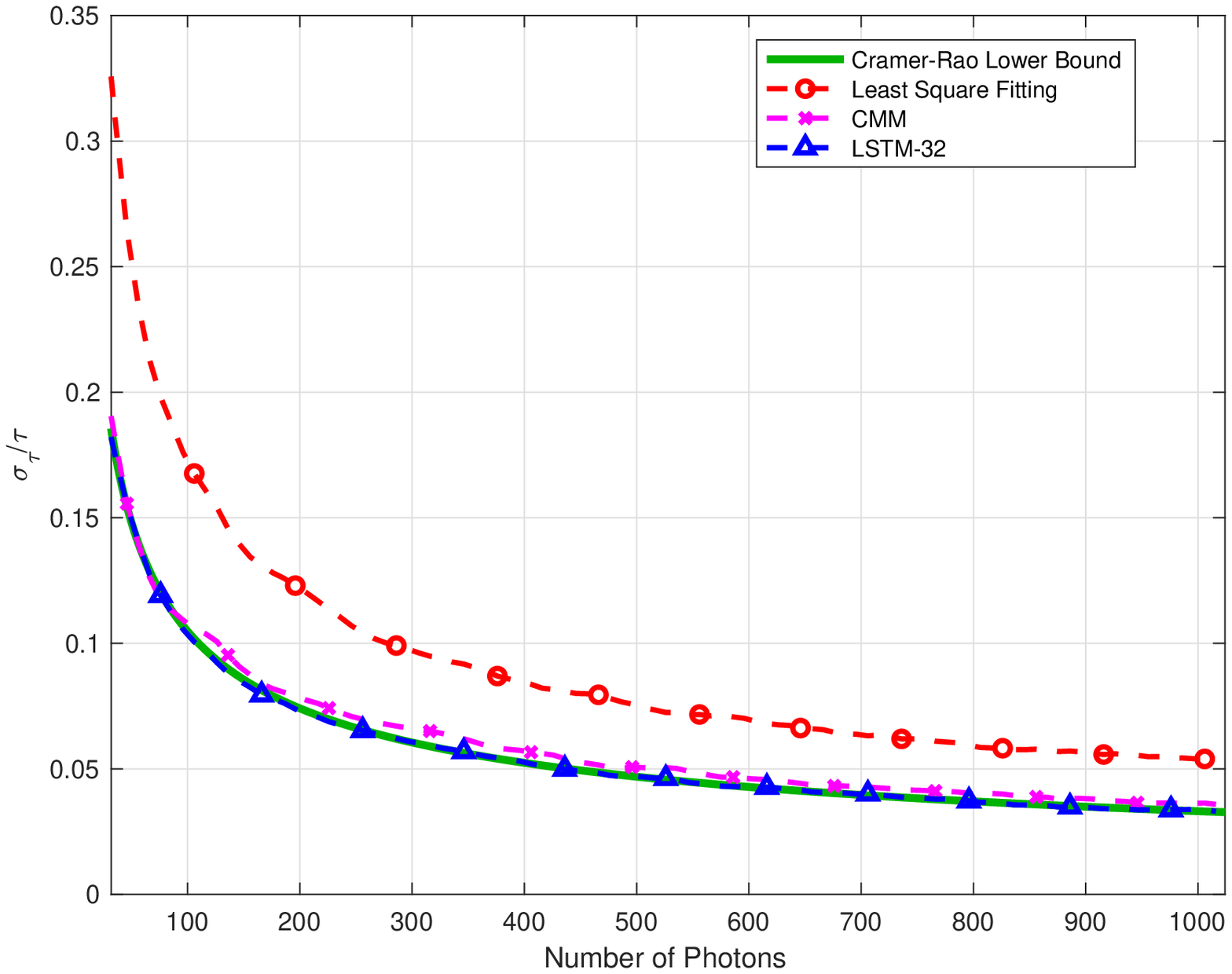}
    }
    % \caption{Cramer-Rao lower bound analysis when including a 1\% background noise level,}
    % \label{fig:crlb-1per}
\end{minipage}

\begin{minipage}{\textwidth}
    \centering
    \subfloat[Relative standard deviation of different estimation methods and CRLB as a function of lifetime for a fixed number of detected photons (1024). 5\% background noise.\label{fig:crlb-lifetime-5per}]{
        \includegraphics[width=0.45\textwidth]{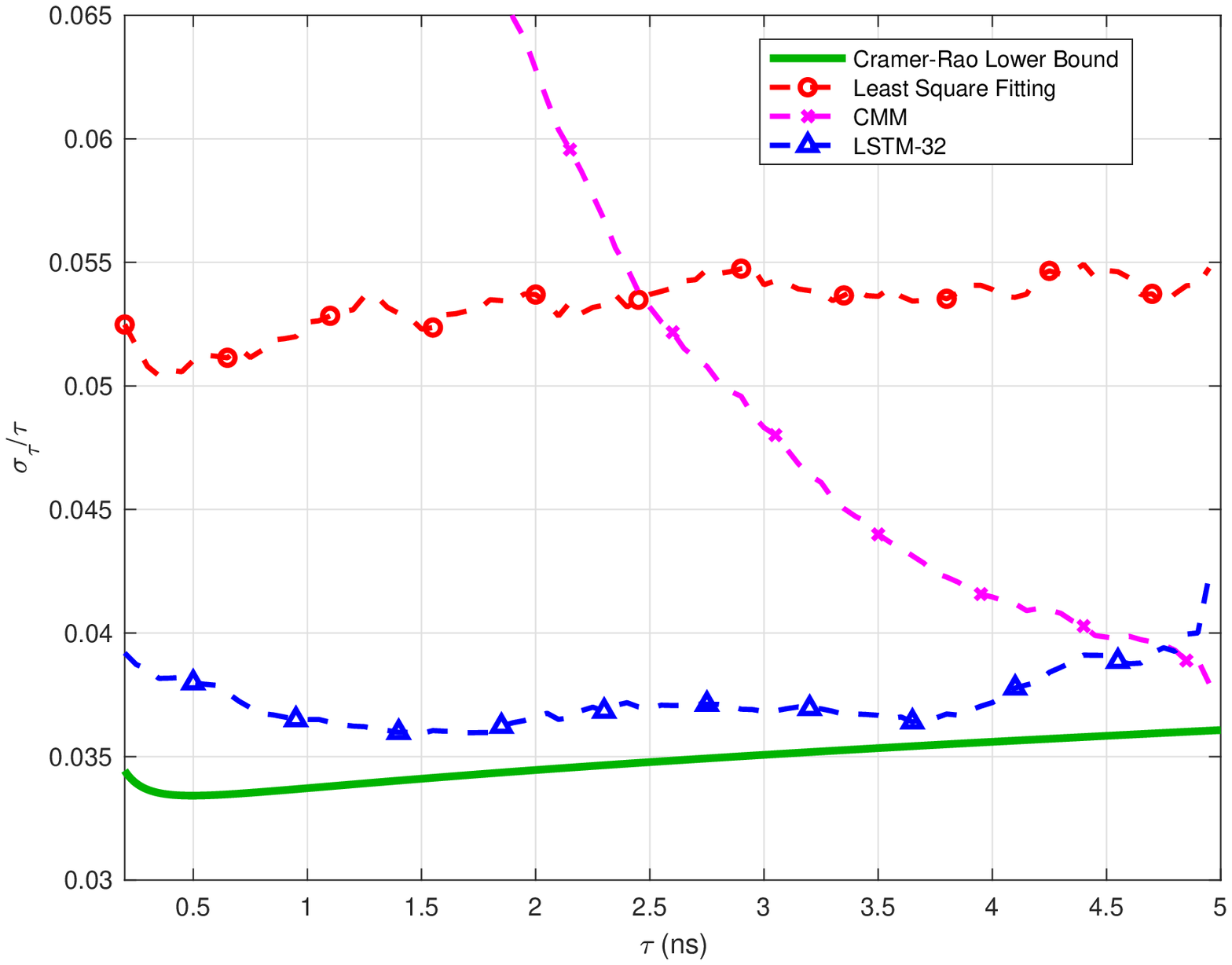}
    }
    \subfloat[Relative standard deviation of different methods and CRLB as a function of the number of detected photons for a fixed lifetime of 2.5 ns. 5\% background noise.\label{fig:crlb-photon-5per}]{
        \includegraphics[width=0.45\textwidth]{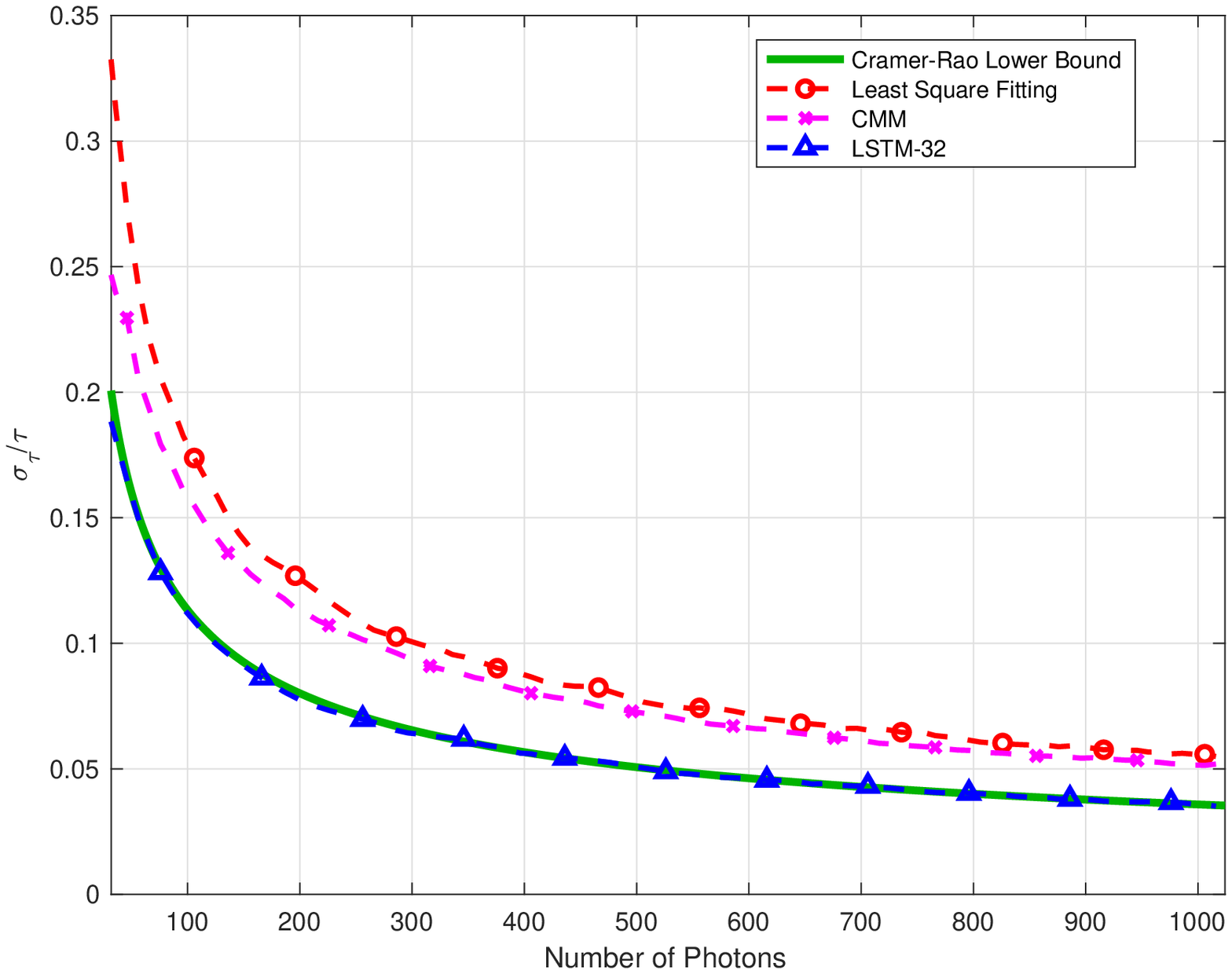}
    }
    % \caption{Cramer-Rao lower bound analysis when including a 5\% background noise level.}
    % \label{fig:crlb-5per}
\end{minipage}

\caption{Cramer-Rao lower bound analysis when including 0\%, 1\%, and 5\% background noise levels.}
\label{fig:crlb-5per}

\end{figure}

To compare the performance with the theoretical optima, the Cramer-Rao lower bound (CRLB) is calculated of the accuracy of the lifetime estimate with an open source software \cite{Bouchet.2019}, given the setting parameters. The variance of the lifetime estimation methods is calculated from Monte Carlo experiments. As for CMM and RNN, 3000 samples are used; as for the least-squares method, 1000 samples are used to reduce running time.

The relationship between lifetime and the relative standard deviation of the different estimators is shown in Figure~\ref{fig:crlb-lifetime}, where the photon count is 1024. One can observe that the variance of CMM and LSTM-32 almost reaches the CRLB, which suggests that CMM and LSTM-32 are near-optimal estimators. Considering that the laser repetition period is much longer than the lifetime and that background noise is not included, it is understandable that CMM reaches the CRLB, since it is approximately a maximum likelihood estimator. LS fitting performs worse than CMM and LSTM-32, which is likely due to the underlying assumption of Gaussian errors. 

The relationship between the number of photons and the relative standard deviation of the different estimators is shown in Figure~\ref{fig:crlb-photon}, where the lifetime is set at 2.5 ns. Similar to Figure~\ref{fig:crlb-lifetime}, the relative standard deviations of CMM and LSTM-32 almost reach the CRLB, while the least square fitting performs worse. This result suggests that CMM and LSTM-32 are efficient estimators over different photon inputs, achieving excellent photon efficiency. They only need less than half of the data to obtain similar results as LS fitting.

We also analyze the CRLB with background noise. The results are shown in Figure~\ref{fig:crlb-5per}. Comparing Figure~\ref{fig:crlb-lifetime-1per} and  Figure~\ref{fig:crlb-lifetime-5per} with Figure~\ref{fig:crlb-lifetime}, we can see the CRLB is lifted a bit in the presence of background noise. The relative standard deviation of LS fitting stays almost unchanged, and that of LSTM-32 increases slightly but is still much better than LS fitting. As for CMM, one can see that the relative standard deviation increases dramatically at shorter lifetimes, which suggests that CMM is very sensitive to background noise for short lifetimes. By comparing Figure~\ref{fig:crlb-photon-1per} and Figure~\ref{fig:crlb-photon-5per} with Figure~\ref{fig:crlb-photon}, we find that the relative standard deviation does not vary with 1\% background noise. With 5\% background noise, however, CMM shows a clear degradation of performance, its relative standard deviation getting close to the one of LS fitting.

\subsection*{Performance on Experimental Dataset}

\begin{figure}
    \centering
    \includegraphics[width=\textwidth]{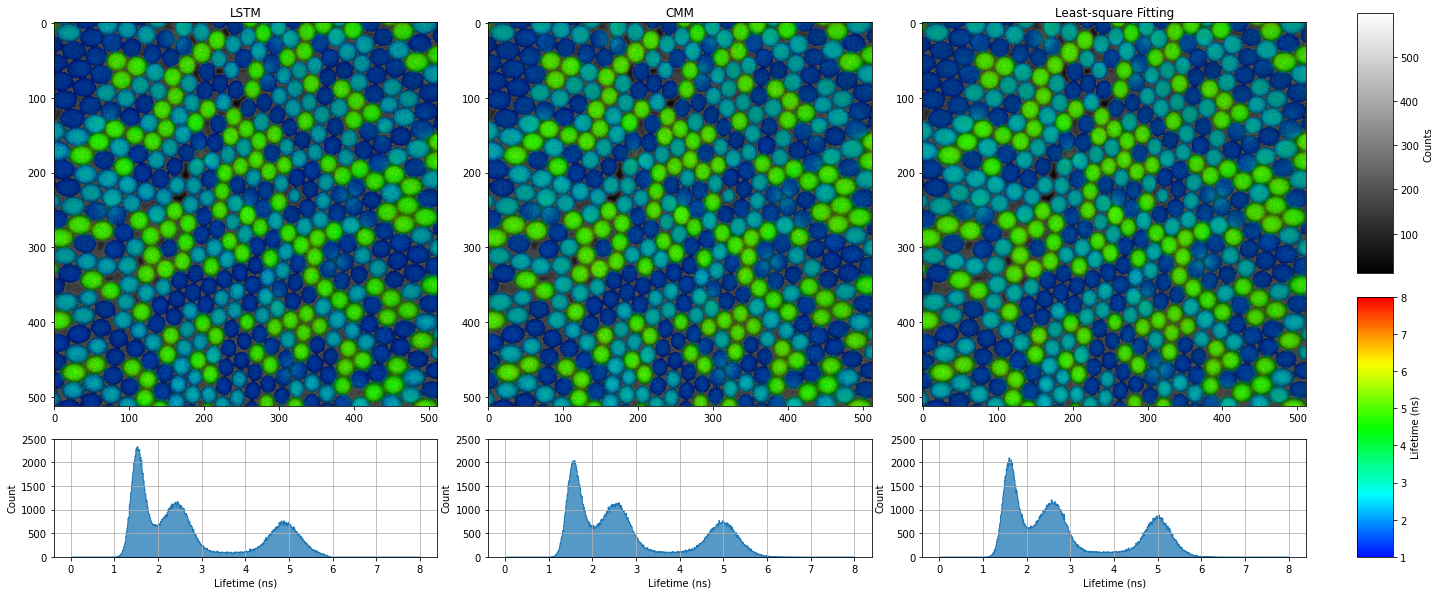}
    \caption{Comparison of LSTM, CMM, and LS Fitting on experimental data. The sample contains a mixture of fluorescent beads with three different lifetimes (1.7, 2.7, and 5.5 ns). The fluorescence lifetime images are displayed using a rainbow scale, where the brightness represents photon counts and the hue represents lifetimes. The lifetime histograms among all pixels are shown below. Most pixels are assumed to contain mono-exponential fluorophores. Two or three lifetimes might be mixed at the edge of the beads.}
    \label{fig:fltbeads}
\end{figure}

To verify the performance of RNNs, which are purely trained on synthetic datasets, on real-world data, the RNNs are tested on experimental data along with CMM and LS fitting as benchmarks. We prepare a fluorescence lifetime-encoded microbeads sample and acquire the TCSPC data with a commercial confocal FLIM setup. It is estimated that the background noise is below 1\%. The LSTM-32 trained with 0\% to 10\% background noise dataset is used. The corresponding results are shown in Figure~\ref{fig:fltbeads}.

The histograms of the three samples share a similar shape. As for LS fitting, an instrument response function (IRF) is estimated from histograms of all pixels and then shared among them, which accounts for its good performance here. The result of CMM has a 2 ns bias, which is corrected by the estimated IRF. It is worth noting that for the first peak in the histogram, LSTM shows a sharper Gaussian shape, which confirms LSTM's good performance under low fluorescence intensity and short lifetime. 

\subsection*{Real-time FLIM Setup with on-FPGA RNN}

\begin{figure}
    \centering
    \includegraphics[width=0.8\textwidth]{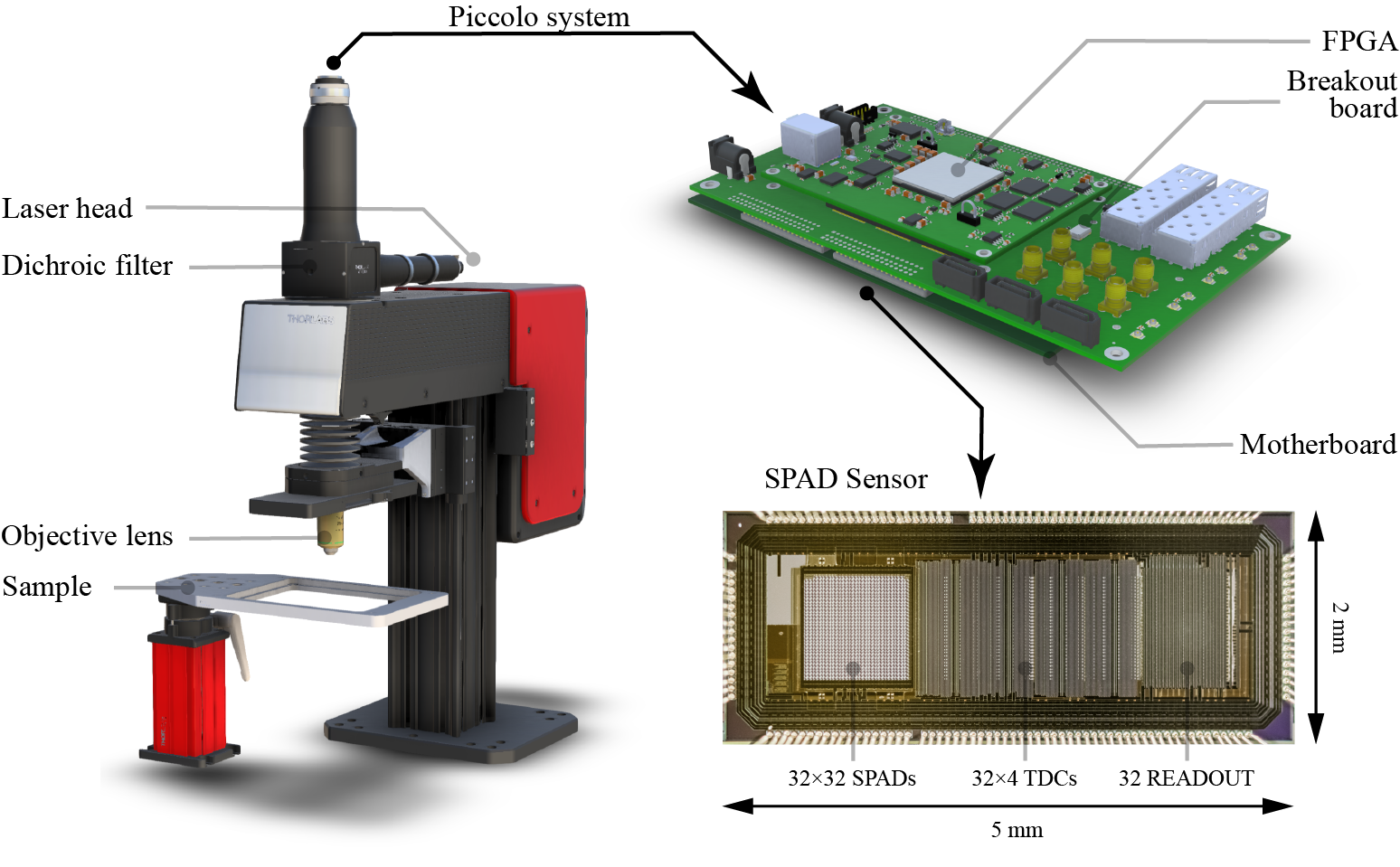}
    \caption{Real-time FLIM system based on the Piccolo 32$\times$32 SPAD sensor and on-FPGA RNNs. The main body of the microscope is from a single-channel Cerna\textsuperscript{\textregistered}  Confocal Microscope System (ThorLabs, Newton, New Jersey, United States). On the top is the Piccolo system, composed of the SPAD sensor itself, motherboard, breakout board, and FPGA. The SPAD sensor has 32$\times$32 SPADs and 128 on-chip TDCs, offering 50 ps temporal resolution. The FPGA is programmed to control the SPAD sensor and communicate with PC through USB 3. The RNN is also deployed on the same FPGA.}
    \label{fig:microscope}
\end{figure}

We further built a real-time FLIM system by utilizing a SPAD array sensor with on-chip time-to-digital converters (TDCs) and deploying the aforementioned RNNs on FPGA for near-sensor processing. The schematic of our setup is shown in Figure~\ref{fig:microscope}. The 32$\times$32 Piccolo SPAD sensor developed at EPFL \cite{Lindner.2018,Zhang.2018} is utilized, on which 128 TDCs offer 50 ps temporal resolution. The sensor is controlled by a Kintex-7 FPGA, where four GRU cores are implemented for lifetime estimation. The four GRU cores are able to process up to 4 million photons per second. While the data transfer rate to the PC is 20Mb/s for histogram mode and 80Mb/s for raw mode, it reduces to only 240kb/s when applying the proposed RNN-based lifetime estimation method.

\begin{figure}
    \centering
    \includegraphics[width=0.7\textwidth]{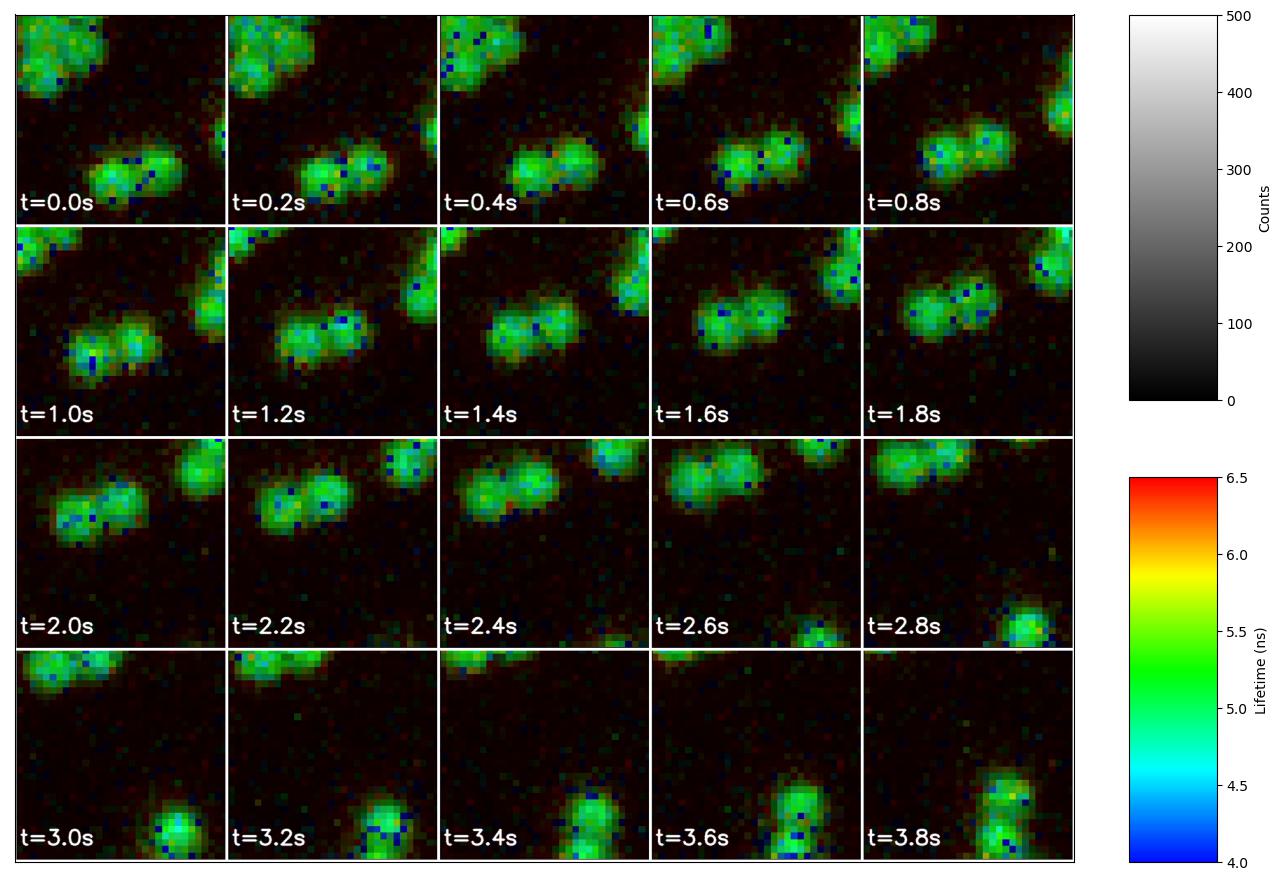}
    \caption{Real-time lifetime image sequence from our FLIM system. The sample contains fluorescent beads with a 5.5 ns reference lifetime. (See the full video in the Supplementary Material)} %%% ADD reference on Supplementary Material VIDEO.
    \label{fig:realtimeflim}
\end{figure}

We prepare a sample containing fluorescent beads with a lifetime of 5.5 ns. The sample is measured by our system in real-time at 5 frames per second. During the imaging, we move the sample plate forward to observe the movement of beads in the images. The result is shown in Figure~\ref{fig:realtimeflim}. The lifetime images are also displayed in rainbow scale. The average photon count for the beads is around 500 per pixel. This illustrates that our system can capture the movement of beads and provide an accurate lifetime estimation. One can also observe that there are some outliers, e.g. dark blue dots and red dots among the green beads. Apart from statistical fluctuations, RNN-based lifetimes tend to be lower when there are not enough photons, which explains why the blue dots are mostly darker than the surrounding pixels. %%% Add lifetime accuracy / bias note...

% The bandwidth of our method is 240Kb/s, histogram mode is 20Mb/s, raw mode is 80Mb/s.

\section*{Discussion}

%In this work, we present an RNN-coupled SPAD TCPSC system for real-time fluorescence lifetime imaging. 
The proposed on-FPGA RNN removes the need for histogramming altogether by taking raw timestamps as input directly, which released hardware resources on FPGA or PC and significantly reduced the burden on data transfer and data processing. The analysis of synthetic data and CRLB shows that RNN, as a data-driven method, reaches excellent accuracy and robustness compared to its competitors, while retaining higher photon efficiency.

The performance of the system can be further improved by using a larger SPAD sensor and by accommodating more RNN cores on the FPGA. A more powerful FPGA or even dedicated neural network accelerators can be used to accommodate more RNN cores. More efficient quantization and approximate schemes can also be explored to reduce resource utilization and latency. In addition, these GRU cores can be further optimized by VHDL implementation. In the future, the RNN cores could be implemented on ASIC and stacked together with SPAD arrays by means of 3-D stacking technology, realizing in-sensor processing\cite{Charbon.129201812122018}.

Though the proposed system is only used for FLI to this date, it can be easily adapted for other applications by retraining the RNN. It can be further combined with other large-scale neural networks for high-level applications, where the output of the RNN is composed of high-level features learned by neural networks automatically, and it serves as input for other neural networks. The existing FLI-based high-level applications such as margin assessment\cite{Unger.2020, Marsden.2021} could also be directly incorporated into our system.

\section*{Methods}

\subsection*{Dataset}

% The dataset is crucial to the success of neural network models. Curating an experimental dataset for training requires extensive labor, which is time-consuming and costly. The specificity of instruments will bring bias to the data and thus impede models' generalization. In this work, we construct synthetic datasets for training and evaluation. A small experimental dataset will also be built for the evaluation of our model on real-world data.

\subsubsection*{Synthetic Dataset}

A simulation that well captures the features of the real scene is the key to constructing synthetic datasets. To accurately model a real FLI system, we take fluorescence decay, instrument response, background noise, and dark counts into account. The latter two are often neglected in previous studies. However, in several scenes such as fluorescence-assisted surgery exist strong background noise which can not be simply ignored. Different from existing NN-based methods, which take histograms as input, we generate synthetic datasets on the timestamp level. Assuming that at most one photon reaches the detector in every repetition period (i.e. the pile-up effect is not considered), the timestamps $t$, namely the arrival time of photons, are modeled as:
\begin{equation}
    t = \sum_{i=1}^{N-1}\mathbf{1}_{k=i}(t_{fluo_i} + t_{irf}) + \mathbf{1}_{k=N} t_{bg},
\end{equation}
where $\mathbf{1}$ is the indicator function, $k$ is the component indicator, $t_{fluo}$ is the fluorescence time delay, $t_{irf}$ is the instrument response time delay, and $t_{bg}$ is the arrival time of background noise or dark counts.

The component indicator $k$ is a random variable with categorical distribution, representing the source of the incoming photon, which can be either a component of the fluorescence decay or background noise. The probability density function (PDF) of $k$ is
\begin{equation}
    f(k|\mathbf{p}) = \prod_{i=1}^{N}p_i^{\mathbf{1}_{k=i}},
\end{equation}
where $p_i$ represents the normalized intensity of fluorescence or background noise.

The fluorescence time delay $t_{fluo_i}$ is subject to an exponential distribution. Its PDF is:
\begin{equation}
    f(t_{fluo_i}|\tau_i) = \frac{1}{\tau_i}e^{-\frac{t_{fluo_i}}{\tau_i}},
\end{equation}
where $\tau_i$ is the lifetime of the fluorescence decay.

The instrument response time delay $t_{irf}$ is subject to a Gaussian distribution. Its PDF is:
\begin{equation}
    f(t_{irf}|t_0, \sigma) = \frac{1}{\sqrt{2\pi}\sigma} e^{-\frac{1}{2}\left(\frac{t_{irf}-t_0}{\sigma}\right)^2},
\end{equation}
where $t_0$ is the peak position, and $\sigma$ can be represented by full width at half maximum(FWHM):
\begin{equation}
    \sigma = \frac{FWHM}{2 \sqrt{2\ln 2}}.
\end{equation}

The time of arrival of background
noise $t_{bg}$ is subject to a uniform distribution. Its PDF is:
\begin{equation}
    f(t_{bg}|T) = \frac{1}{T},
\end{equation}
where $T$ is the repetition period.

Given a set of the above parameters, synthetic datasets can be generated with different lifetime ranges, background noise ratios, and components of fluorescence. In this work, the FWHM is assumed to be 167.3 ps, in accordance with previous studies\cite{Xiao.2021, Zang.3252022}; $t_0$ for each sample is generated from a uniform distribution from 0 to 5 ns. To train models in the presence of background noise, $p_N$ for each sample is generated from a uniform distribution from 0 to $10\%$. Each dataset contains 500,000 samples, and each sample contains 1024 timestamps.

\subsubsection*{Experimental Dataset}

Testing the model, which is purely trained on synthetic data, on experimental data is essential to ensure its applicability in real-world scenarios, thus an experimental dataset is curated for evaluation. Fluorescent beads from PolyAn with reference lifetimes of 1.7, 2.7, and 5.5 ns are adopted as sample. The beads are made of 3D-carboxy, with a diameter of 6.5 $\mu$m. The excitation wavelength is around 488 nm and the emission spectra are 602-800 nm, 545-800 nm, and 559-718 nm, respectively. The fluorescence intensity of these three beads is around 1:2:5. Fluorescent beads with different lifetimes are mixed together with all possible combinations, and put in a 384-well plate for imaging.

A commercial FLIM system, available at the Bioimaging and Optics Platform (BIOP) of EPFL, is utilized to measure the sample and acquire the experimental data. A confocal microscope (Leica SP 8, w/ HyD SMD detector) is used for imaging, a super-continuum laser (NKT Photonics, SuperK Extreme EXW-45) is used for illumination, and a TCSPC module (PicoHarp 300 TCSPC) is used for time-tagging. The sample is excited under a 20 MHz laser, corresponding to a repetition period of 50 ns. The excitation wavelength is 486 nm and the spectrum of the emission filter ranges from 600 to 700 nm. The temporal resolution of time-tagging is 16 ps.

\subsection*{Neural Network}

The neural network is first built, trained, and evaluated on the PC with PyTorch\cite{paszke2019pytorch}. Then its weights are quantized and the activation functions are approximated. After that, the neural network is written in C/C++, loading the quantized weights and approximated activation functions, and is further translated into hardware description language (HDL) by Vitis High-level Synthesis (HLS).

\subsubsection*{Model}

Three RNN variants are adopted here, which are simple RNN, GRU, and LSTM. The default models in PyTorch are used, whereas the input size is 1, so selected to accommodate the timestamps. Considering the hardware limitation, only single-layer RNNs are considered. The hidden sizes range from 8 to 64. Since the timestamps are processed in real-time and are not stored, bidirectional RNNs cannot be used. An FCNN with one hidden layer takes the hidden state as input to predicts the lifetime.

\subsubsection*{Training}

Normally, the loss function for RNNs is built on the output of the last timestep or the average output of all timesteps. In fluorescence lifetime estimation, the performance of estimators is supposed to be improved with more photons. Under this principle, we design a weighted mean square percentage error (MSPE) function, assigning more importance to subsequent timesteps:
\begin{equation}
    L(\mathbf{y}, \mathbf{\hat{y}}) = \sum_{i=1}^{N} w_i  \left( \frac{y_i - \hat{y}_i}{y_i}\right)^2,
\end{equation}
where $N$ is the number of timesteps, $\mathbf{y}$ is the ground truth, $\mathbf{\hat{y}}$ the prediction, and $w_i$ the weight at timestep $i$:
\begin{equation}
    w_i = \frac{1}{1 + e^{-\left( \frac{i - N/4}{N/4} \right)}}.
\end{equation}

The weights for hidden states are initialized by an orthogonal matrix. All biases are initialized with 0s. For LSTM, the weights for cell states are initialized by Xavier initialization \cite{glorot2010understanding}, and the bias for forget gates is initialized with 1s.

The dataset is randomly split into training, evaluation, and test set, with the ratio of sizes being 8:1:1. The batch size is 32. Adam optimizer is used with an initial learning rate of 0.001\cite{kingma2014adam}. The learning rate decays every 5 epochs at the rate of 0.9. The whole training process takes 100 epochs.

\subsubsection*{Evaluation}

Three metrics are used to evaluate the performance of RNNs and benchmarks on synthetic data, which are:

\begin{equation}
    \mathrm{RMSE} = \frac{\sqrt{\sum_{i=1}^{N}{(y_i - \hat{y}_i)^2}}}{N},
\end{equation}

\begin{equation}
    \mathrm{MAE} = \frac{\sum_{i=1}^{N} |y_i - \hat{y}_i|}{N},
\end{equation}

\begin{equation}
    \mathrm{MAPE} = \frac{\sum_{i=1}^{N} |\frac{y_i - \hat{y}_i}{y_i}|}{N}.
\end{equation}

% \begin{table}[!tbp]
%     \centering
%     \begin{tabular}{cccc}
%     \toprule
%         Metrics & RMSE & MAE & MAPE \\
%     \midrule
%        Definition  & $\frac{\sqrt{\sum_{i=1}^{N}{(y_i - \hat{y}_i)^2}}}{N}$ & $\frac{\sum_{i=1}^{N} |y_i - \hat{y}_i|}{N}$ & $\frac{\sum_{i=1}^{N} |\frac{y_i - \hat{y}_i}{y_i}|}{N}$ \\
%     \bottomrule
%     \end{tabular}
%     \caption{Metrics used for evaluation. $y$ is the ground truth, $\hat{y}$ is the prediction, and $N$ is the number of samples.}
%     \label{tab:metrics}
% \end{table}

\subsubsection*{Cramer-Rao Lower Bound}

Cramer-Rao lower bound (CRLB) gives the best precision that can be achieved in the estimation of fluorescence lifetime\cite{Kollner.1992,Kim.2013,Bouchet.2019}. Mathematically, CRLB expresses a lower bound of variance of estimators and it is proportional to the inverse of the Fisher information:

\begin{equation}
	Var(\hat{\theta}) \ge \frac{(f'(x; \theta))^2}{\mathcal{J} (\theta)},
\end{equation}

where $f(x; \theta))$ is the PDF and $\mathcal{J}$ is the Fisher Information, which is defined as:

\begin{equation}
	\mathcal{J} (\theta) = nE_{\theta}\left[ \left( \frac{\partial}{\partial \theta} \ln f(x; \theta)  \right) ^2 \right].
\end{equation}

The CRLB is calculated with open-source software\cite{Bouchet.2019}.

\subsubsection*{FPGA Implementation}

Quantization is an effective way to reduce resource utilization and latency on hardware. In common deep learning frameworks, such as PyTorch or Tensorflow, model weights and activations are represented by 32-bit floating point numbers. However, it would be inefficient to perform operations for floating point numbers with such bitwidth. We aim to quantize the 32-bit floating point numbers with fixed-point numbers and to reduce the bitwidth as much as possible, while maintaining the same model behavior.

Both PyTorch and TensorFlow provide tools of quantization for edge devices, namely PyTorch Quantization and TensorFlow Lite. However, the quantized models rely on their own libraries to run, and the quantized weights cannot be exported. Therefore, we use Python and an open-source fixed point number library to realize a quantized GRU for evaluation. We compare 8-bit, 16-bit, and 32-bit fixed-point numbers to quantize weights and activations separately. The results show that the weights can be quantized to 16-bit fixed point numbers without a significant accuracy drop, and to 8-bit fixed point numbers with an acceptable accuracy drop. Activations can be quantized to 16-bit fixed point numbers without a significant accuracy drop, but 8-bit fixed point quantization will lead the model to collapse. Besides the fixed point precision, we find that the rounding method has a great impact on the performance. Truncating, often the default rounding method, brings larger error. Fixed point numbers with convergent rounding have almost the same behavior as floating point numbers.

The quantized GRU model is then implemented on FPGA. For convenience, the GRU is written in C++ and compiled to Vivado IP with Vitis HLS. The whole model is divided into two parts: a GRU core and an FCNN. The GRU core is designed to be shared among a group of pixels, and the FCNN will be run sequentially for each pixel after integration. Upon receiving a timestamp, GRU core loads hidden states from block RAMs (BRAMs), updates the hidden states, and sends them back to BRAM. After the integration of each repetition period, the FCNN loads the hidden state from BRAM, and streams the estimated lifetime to a FIFO.

% For the implementation on FPGA, an LSTM with 12 hidden units is trained and tested. It only has 672 parameters and keeps the MAPE below 4\%. The activation functions are approximated by piecewise linear functions to reduce latency and resource utilization. 32-bit fixed point quantization for both weights and activations is used to preserve the highest precision here, with 5-bit integer precision and 27-bit fraction precision. The interface of the LSTM cell includes a 42-bit unsigned integer event signal, and 2 BRAM interfaces for hidden and cell states respectively. The interface of the FCNN includes a BRAM interface for the hidden state and a FIFO interface for output streaming. 12 single-port BRAMs are used to store the cell state, and 12 dual-port BRAMs are used to store the hidden state.

\subsection*{Experimental Setup}

A real-time FLI microscopy (FLIM) system with our SPAD sensor and on-FPGA RNN is built, which is shown in Figure.~\ref{fig:microscope}. The microscope is adapted from the sa confocal microscope system (Single-Channel Cerna\textsuperscript{\textregistered}  Confocal Microscope System), though it is only used for widefield imaging in this work. The same fluorescent bead samples are measured, hence a 480 nm pulsed laser (PicoQuant) is utilized. A set of filters is adopted for fluorescence imaging. The excitation filter (Thorlabs FITC Excitation Filter) has a central wavelength of 475 nm with a bandwidth of 35 nm. The emission filter is a long-pass filter (Thorlabs Ø25.0 mm Premium Longpass Filter) with a cut-on wavelength of 600 nm. The dichroic filter (Thorlabs GFP Dichroic Filter) has a reflection band from 452 to 490 nm and a transmission band from 505 nm to 800 nm.

The Piccolo system is used for single-photon detection and time tagging\cite{Zhang.2018}. The complete system, along with its components and a micrograph of the Piccolo chip is shown in Figure~\ref{fig:microscope}. Piccolo provides 50-ps temporal resolution and 47.8\% peak photon detection probability (PDP). Versions with microlenses are available as well, to improve the light collection efficiency. The median dark count rate (DCR) is 113  cps (per pixel at room temperature). A Xilinx FPGA was used to communicate with the PC and control the sensor. To minimize the system and reduce latency, the RNNs were deployed on the same FPGA.

\tikzstyle{rect} = [rectangle, rounded corners, minimum width=2cm, minimum height=0.5cm,text centered, draw=black]

\tikzset{
    position/.style args={#1:#2 from #3}{
        at=(#3.#1), anchor=#1+180, shift=(#1:#2)
    }
}

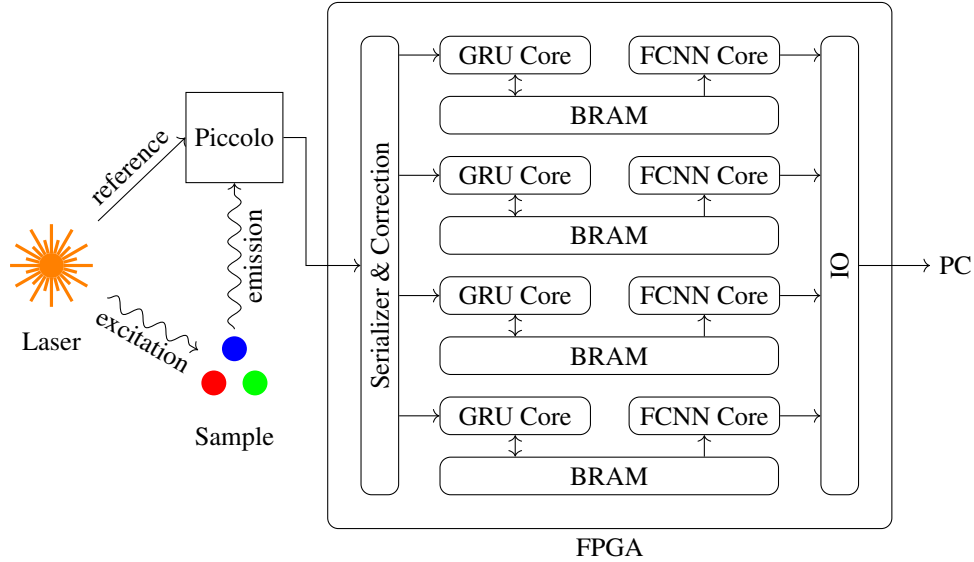
\begin{figure}
    \centering
    \begin{tikzpicture}[node distance=2cm]

    \foreach \i in {0,1,...,3}{
        \node (grucore\i) [rect]  at (0, -1.6*\i){GRU Core};
        \node (fcnncore\i) [rect, right= 0.5cm of grucore\i] {FCNN Core};
        \node (bram\i) [rect, below = 0.8cm of grucore\i.west, anchor=west, minimum width=4.5cm] {BRAM};
    
        \draw [<->] (grucore\i) -- (grucore\i|-bram\i.north);
        \draw[<-] (fcnncore\i) -- (fcnncore\i|-bram\i.north);
    }

    \node(io) [rect, rotate=90, minimum width=6.1cm, minimum height=0.5cm, right=1.8cm of fcnncore0.north, anchor=east] {IO};

    \node(serializer) [rect, rotate=90, minimum width=6.1cm, minimum height=0.5cm, left=1.8cm of grucore0.north, anchor=east] {Serializer \& Correction};

    \foreach \i in {0,1,...,3}{
        \draw [<-] (grucore\i) -- (grucore\i-|serializer.south);
        \draw[->] (fcnncore\i) -- (fcnncore\i-|io.north);
    }

    \node [rect,anchor=center,minimum height=7cm,minimum width=7.5cm,label=below:FPGA] at (current bounding box.center)(fpga) {};

    \node [right=0.5cm of fpga] (pc) {PC};
    \draw [->] (io) -- (pc);

    % light source
    \node [outer sep=0pt,circle, fill=orange, left=3.5cm of fpga, minimum width=0.3cm, minimum height=0.3cm,label={[below,label distance=0.6cm]-90:{Laser}}] (laser) {};

    \foreach \i in {0,1,...,11}{
    \draw [very thick,orange] (laser) -- ++(30*\i:0.55cm);
    \draw [very thick,orange] (laser) -- ++(15+30*\i:0.35cm);
    }
    % sample
    \node [position=-30:{2.5cm} from laser, label={[below,label distance=0.5cm]-90:{Sample}}] (samplecenter) {};
    \node [circle, fill=green, position=-30:{0cm} from samplecenter] {};
    \node [circle, fill=blue, position=90:{0cm} from samplecenter] {};
    \node [circle, fill=red, position=210:{0cm} from samplecenter] {};

    % piccolo
    \node [rectangle,draw, minimum width=1.2cm,minimum height=1.2cm,above=2.5cm of samplecenter.center] (piccolo) {Piccolo};

    % placeholder
    \node [circle, minimum width=1.7cm] (laserph) at (laser.center) {};
    \node [circle, minimum width=1.1cm] (sampleph) at (samplecenter.center) {};

    \draw [->, decorate, decoration={snake}] (laserph) -- node[below,sloped] {excitation} (sampleph);
    \draw [->, decorate, decoration={snake}] (sampleph) -- node[below,sloped] {emission} (piccolo);
    \draw [->] (laserph) -- node[above,sloped] {reference} (piccolo.west);
    \draw [->] (piccolo.east) -- ++ (0.3,0) |- (serializer.north);
    
    \end{tikzpicture}
    \caption{Schematic of the proposed real-time FLIM system with on-FPGA GRUs. A pulsed laser illuminates the sample repeatedly and sends a reference signal to Piccolo to reset TDCs at the same time. The emitted fluorescence photon is then detected by Piccolo and time-tagged, whose arrival time is sent to FPGA in parallel for further processing. The incoming timestamps are serialized and corrected and then sent to GRU cores. The hidden states of GRU are stored in the BRAM. Upon the arrival of a timestamp, the hidden state of the corresponding pixel is read by the GRU core, then the hidden state is updated and written back to the BRAM. After the integration period, the final hidden states are read by FCNN cores and the lifetime is estimated and sent to PC.}
    \label{fig:fpga}
\end{figure}

The schematic of the FPGA design is shown in Figure~\ref{fig:fpga}. Four computation units are realized, each of which is in charge of a quarter of the sensor (32 by 8 pixels). The timestamps, sent to FPGA in parallel, are serialized and distributed to four computation units based on their SPAD IDs. Each computation unit is composed of one GRU core, one two-layer fully connected neural network (FCNN) core, and one BRAM. The computation speed is mainly limited by the latency of the GRU core, which is 1.05 ns when employing a 160 MHz clock. The photons that arrive when computation units are busy are simply discarded. The four computation units together are capable of processing up to 4 million photons per second.

\bibliographystyle{unsrt}
\bibliography{sample}  %%% Uncomment this line and comment out the ``thebibliography'' section below to use the external .bib file (using bibtex) .




% 	\newblock {\em arXiv preprint arXiv:1804.09028}            , 2018.

% \end{thebibliography}

\end{document}